\documentclass[journal]{IEEEtran}
\usepackage{threeparttable}
\usepackage{fancyhdr}
\usepackage{amsmath}
\usepackage{color}
\usepackage{soul}
\usepackage{mathrsfs}
\usepackage{amssymb}
\usepackage{graphicx}
\usepackage{amsfonts}
\usepackage{booktabs}
\usepackage{algorithm} 
\usepackage{algorithmic} 
\usepackage{extarrows}
\begin{document}
\bibliographystyle{IEEEtran}
\title{Joint Channel Training and Feedback for FDD Massive MIMO Systems}
\author{\IEEEauthorblockN{Wenqian Shen, Linglong Dai, Yi Shi, Byonghyo Shim, and Zhaocheng Wang}

\thanks{Copyright (c) 2015 IEEE. Personal use of this material is permitted. However, permission to use this material for any other purposes must be obtained from the IEEE by sending a request to pubs-permissions@ieee.org.}
\thanks{W. Shen, L. Dai, and Z. Wang are with the Department of Electronic Engineering, Tsinghua University, Beijing 100084, China (e-mail: swq13@mails.tsinghua.edu.cn; \{daill, zcwang\}@tsinghua.edu.cn).}
\thanks{Y. Shi is with Huawei Technologies, Beijing 100095, China (e-mail: wn.shiyi@gmail.com).}
\thanks{B. Shim is with Institute of New Media and Communications, School of Electrical and Computer Engineering, Seoul National University, Seoul 151-742, Korea (e-mail: bshim@snu.ac.kr).}
\thanks{This work was supported by the National Key Basic Research Program of China (Grant No. 2013CB329203),  the National Natural Science
Foundation of China (Grant Nos. 61571270 and 61201185),  the Beijing Natural Science Foundation (Grant No. 4142027), and the Foundation of Shenzhen government.}}

\maketitle
\begin{abstract}
Massive multiple-input multiple-output (MIMO) is widely recognized as a promising technology for future 5G wireless communication systems. To achieve the theoretical performance gains in massive MIMO systems, accurate channel state information at the transmitter (CSIT) is crucial. Due to the overwhelming pilot signaling and channel feedback overhead, however, conventional downlink channel estimation and uplink channel feedback schemes might not be suitable for frequency-division duplexing (FDD) massive MIMO systems. In addition, these two topics are usually separately considered in the literature. In this paper, we propose a joint channel training and feedback scheme for FDD massive MIMO systems. Specifically, we firstly exploit the temporal correlation of time-varying channels to propose a differential channel training and feedback scheme, which simultaneously reduces the overhead for downlink training and uplink feedback. We next propose a structured compressive sampling matching pursuit (S-CoSaMP) algorithm to acquire a reliable CSIT by exploiting the structured sparsity of wireless MIMO channels. Simulation results demonstrate that the proposed scheme can achieve substantial reduction in the training and feedback overhead.
\end{abstract}

\IEEEpeerreviewmaketitle

\section{Introduction}
Massive multiple-input multiple-output (MIMO) is widely recognized as a key technology for future 5G wireless communications due to its substantial gains in spectrum and energy efficiency.
In massive MIMO systems, the base station (BS) is equipped with a large number of antennas to provide high spatial degrees of freedom (DoF).
To fully capitalize the DoF gain provided by massive MIMO systems, channel state information at the transmitter (CSIT) is essential.
Recall that CSIT is crucial in channel adaptive techniques such as beamforming, power allocation, and interference alignment \cite{ScalingupMIMO}.
In time-division duplexing (TDD) systems, channel reciprocity can be used to obtain the downlink CSI via uplink channel estimation. However, this feature cannot be exploited in frequency division duplexing (FDD) systems.
Since FDD systems are more widely deployed and show some advantages over TDD systems in delay-sensitive and traffic-symmetric applications \cite{Shim},
it is of great importance to study downlink channel training and uplink channel feedback in FDD massive MIMO systems.

For conventional channel training and feedback scheme, the BS firstly transmits orthogonal pilots, the number of which scales linearly with the number of BS antennas.
Then, users estimate the CSI using the conventional technique such as least squares (LS) or minimum mean square error (MMSE) algorithm \cite{Ozdemir}.
The estimated CSI is then fed back to the BS via dedicated uplink resources.
In massive MIMO systems, the number of BS antennas might be an order of magnitude larger than that of state-of-the-art LTE-Advanced system, so the training overhead is a serious concern \cite{Dai14}.
Furthermore, the massively expanded MIMO channel matrix also renders precise CSI feedback an extremely challenging problem \cite{Shim}.
To alleviate the overhead, several downlink pilot design and uplink CSI feedback schemes have been proposed \cite{Noh_et_al,Choi_et_al,Samardzija_and_Mandayam,Caire_et_al,ZhuXD,HanYJ,protocals,IET}.
Among these, compressive sensing (CS) has merged as a promising technique in recent years.
The fundamental principle of CS-based schemes relies on the fact that broadband wireless channels have sparse channel impulse response (CIR) due to the limited number of significant paths \cite{Dai14,ZhuXD,HanYJ}.
To date, number of CS-based channel estimation schemes have been proposed where the sparse channel impulse response (CIR) can be recovered from a reduced number of received pilots \cite{ZhuXD,HanYJ}.
Nevertheless, the reduction of training overhead is insignificant when CIR is not sparse enough.
On the other hand, several CS-based channel feedback schemes have also been proposed.
The underlying assumption on this work is that users firstly achieve perfect CSI estimation, and then compressing (through projection) the sparse (or compressible in some transform domains) CSI into low-dimensional measurements for feedback \cite{protocals,IET}.
However, the separate treatments of the two coupled communication procedures (downlink channel training and uplink channel feedback) result in some performance loss, and induce unnecessary computational overhead especially for power-limited users.
In \cite{RaoJ}, an approach based on compressive sensing (CS) has been proposed to reduce both the downlink training and uplink CSI feedback overhead.
Although this approach is promising when the channel matirces of different users are sparse and partially share common support, it is not effective when these assumptions are violated.

In this paper, we propose a structured-CS based differential joint channel training and feedback scheme for massive MIMO systems,
where downlink training and uplink feedback are considered in a joint manner.
We design a system such that users will directly feed back the received pilots to the BS without the channel estimation.
BS estimates the CSI using CS-based algorithm.
Such ``closed-loop" operation streamlines the CSIT acquisition process by removing channel estimation and the compression procedures at the user side.
In addition, our proposed scheme also features a differential operation, which exploits the temporal correlation between two consecutive CIRs to reduce the number of required pilots.
Finally, due to the close antenna spacing at the BS and resulting similar path delays, CIRs associated with different BS antennas usually have a common support (i.e. the locations of non-zero elements) \cite{CommonSupport}. By exploiting this feature, we propose a structured compressive sampling matching pursuit (S-CoSaMP) algorithm to further reduce the training as well as feedback overhead.

The remainder of this paper is organized as follows. Section II briefly introduces the massive MIMO system model, where the temporal and spatial correlations of channels are emphasized. Section III addresses the proposed differential joint channel training and feedback scheme, together with the proposed S-CoSaMP algorithm. Section IV presents the simulation results. Finally, conclusions are drawn in Section V.

Notations: Lower-case and upper-case boldface letters denote vectors and matrices, respectively;
$(\cdot)^T$, $(\cdot)^H$ and $(\cdot)^{-1}$ denote the transpose, conjugate transpose and inverse of a matrix, respectively;
$\mathbf{\Theta}^{\dagger}=(\mathbf{\Theta}^H\mathbf{\Theta})^{-1}\mathbf{\Theta}^{H}$ is the Moore-Penrose pseudoinverse of $\mathbf{\Theta}$;
$\mathbf{\Theta}_S$ denotes the sub-matrix consisted of columns of $\mathbf{\Theta}$ according to the index set $S$.
$\|\cdot\|_p$ is the $l_p$-$ \text{norm}$;
$S^c$ denotes the complementary set of $S$;
$\mathcal{T}(\mathbf{x},K)$ denotes a prune operator on $\mathbf{x}$ that sets all but $K$ elements with the largest amplitudes to zero;
$\Gamma_\mathbf{x}$ denotes the support of $\mathbf{x}$, i.e., $\Gamma_\mathbf{x}=\{i,~\mathbf{x}(i)\neq0\}$.
\section{System Model}
\subsection{System Model}
We consider a massive MIMO system operating in FDD mode with the ubiquitous orthogonal frequency division multiplexing (OFDM) modulation.
There are $M$ antennas at the BS and $U$ scheduled single-antenna users.
The length of one OFDM symbol is $N$.
The BS transmits pilots $\mathbf{c}_i\in\mathcal{C}^{P\times1}$ at the $i$-th transmit antenna, where $i=1,2\cdots,M$, and $P$ is the number of pilots.
At a certain user, the received pilots $\mathbf{y}_\Omega$ in the frequency domain can be expressed as
\begin{equation}
\mathbf{y}_\Omega=\sum_{i=1}^{M}\mathbf{C}_i(\mathbf{F}_L)_\Omega\mathbf{h}_i+\mathbf{n}_\Omega,
\end{equation}
where $\mathbf{C}_i=\text{diag}\{\mathbf{c}_i\}$,
$\mathbf{F}_L\in\mathcal{C}^{N\times L}$ is a sub-matrix consisting of the first $L$ columns of the discrete fourier transform (DFT) matrix of size $N\times N$,
$(\mathbf{F}_L)_\Omega$ is the sub-matrix consisting of the rows of $\mathbf{F}_L$ with indices from the index set $\Omega$ of subcarriers assigned to pilots,
which can be randomly selected from the subcarrier set $\{1,2,\cdots,N\}$,
$\mathbf{h}_i=[\mathbf{h}_i(1), \mathbf{h}_i(2),\cdots, \mathbf{h}_i(L)]^T$ is the CIR between the $i$-th BS antenna and the user.
Due to the physical propagation characteristics of multi-path channels \cite{Dai14,ZhuXD,HanYJ},
it is assumed that the number of non-zero elements $K$ in CIR $\mathbf{h}_i$ is much smaller than the maximal channel length $L$.
This vector is often referred to as $K$-sparse vector.
The parameter $\mathbf{n}_\Omega=[n_1, \cdots, n_P]^T$ represents the independent and identically distributed (i.i.d.) additive white complex Gaussian noise (AWGN) with zero mean and variance $\sigma_n^2$ imposed on pilots.
For notation simplicity, the equation (1) can also be written as
\begin{equation}
\mathbf{y}_\Omega=\mathbf{\Theta}\mathbf{h}+\mathbf{n}_\Omega,
\label{eq2}
\end{equation}
where $\mathbf{\Theta}=[\mathbf{C}_1(\mathbf{F}_L)_\Omega,\mathbf{C}_2(\mathbf{F}_L)_\Omega,\cdots, \mathbf{C}_{M}(\mathbf{F}_L)_\Omega]$, and $\mathbf{h}=[\mathbf{h}_1^T,\mathbf{h}_2^T,\cdots, \mathbf{h}_{M}^T]^T$ denotes the aggregate CIR from $M$ BS antennas.
In the sequel, we refer to the aggregate CIR as CIR unless it causes ambiguity.
\subsection{Temporal Correlation of Time-Varying MIMO Channels}
We consider block-fading MIMO channels, where the CIR $\mathbf{h}$ changes from slot to slot but remains unchanged during one time slot.
The CIR series $\{\mathbf{h}^{(t)}\}_{t=0}^{T-1}$ in $T$ consecutive time slots usually exhibits temporal correlation, and thus is compressible in the time domain \cite{protocals,Dynamic}.
The dynamic channel can be modeled by the variation of CIR's support and the evolution of the non-zero elements' amplitudes as \cite{ZhuXD,HanYJ}
\begin{equation}
  \mathbf{h}^{(t)}=\mathbf{s}^{(t)}\circ\mathbf{a}^{(t)},
  \label{eq3}
\end{equation}
where $\mathbf{s}^{(t)}(l)\in \{0,1\}$, $\mathbf{a}^{(t)}(l)\in\mathcal{C}$, and $\circ$ denotes the Hadamard product.
The variation of $\{\mathbf{s}^{(t)}(l)\}_{t=0}^{T-1}$ over time can be modeled as a first-order Markov process \cite{Markov},
which can be fully characterized by two transition probabilities $p_{10}=\text{Pr}\{\mathbf{s}^{(t+1)}(l)=1|\mathbf{s}^{(t)}(l)=0\}$ and $p_{01}=\text{Pr}\{\mathbf{s}^{(t+1)}(l)=0|\mathbf{s}^{(t)}(l)=1\}$,
and a distribution $\mu^{(0)}=\text{Pr}\{\mathbf{s}^{(0)}(l)=1\}$ in the initial time slot $t=0$.
For the steady-state Markov process, where $\text{Pr}\{\mathbf{s}^{(t)}(l)=1\}=\mu$, for all $t$ and $l$, only two parameters $p_{01}$ and $\mu$ are sufficient to characterize the process since $p_{10}=\mu p_{01}/(1-\mu)$.
The evolution of amplitude $\{\mathbf{a}^{(t)}(l)\}_{t=0}^{T-1}$ over time can be modeled by the first-order autoregressive model as \cite{AR}
 \begin{align}
  \mathbf{a}^{(t)}(l)=\rho\mathbf{a}^{(t-1)}(l)+\sqrt{1-\rho^2}\mathbf{w}^{(t)}(l),
    \label{eq4}
\end{align}
where the correlation coefficient $\rho =J_0(2\pi f_d\tau)$ is given by the zero-order Bessel function of the first kind with $f_d$ being the maximal Doppler frequency and $\tau$ being the time slot duration,
the parameter $\mathbf{w}^{(t)}(l)\sim\mathcal{CN}(0,\sigma_{\omega}^2)$ is the i.i.d. complex Gaussian variables.

\subsection{Spatial Correlation of Massive MIMO Channels}
Due to the close antenna spacing at the BS,
CIRs $\{\mathbf{h}_i^{(t)}\}_{i=1}^M$ between BS antennas and the single receive antenna of a user have similar path delays.
Thus, they share a common support \cite{CommonSupport}, i.e.,
\begin{equation}
\Gamma_{\mathbf{h}_1^{(t)}}=\Gamma_{\mathbf{h}_2^{(t)}}=\cdots =\Gamma_{\mathbf{h}_{M}^{(t)}},
\end{equation}
This property of $\mathbf{h}^{(t)}$ is referred to as structured sparsity.
Thus, $\{\mathbf{h}_i^{(t)}\}_{i=1}^M$ can be generated with the same support vectors $\{\mathbf{s}_i^{(t)}\}_{i=1}^M$, where $\{\mathbf{s}_i^{(t)}\}_{i=1}^M$ can be generated at random without loss of generality.
\section{Differential Joint Channel Training and Feedback Based on S-CoSaMP}
In this section, we present the proposed differential joint channel training and feedback scheme. By exploiting the structured sparsity of massive MIMO channels, we propose the S-CoSaMP algorithm to recover CIR $\mathbf{h}^{(t)}$ from received pilots $\mathbf{y}_\Omega^{(t)}$ at the BS.
\subsection{Differential Joint Channel Training and Feedback}
\begin{figure}
\center{\includegraphics[width=0.5\textwidth]{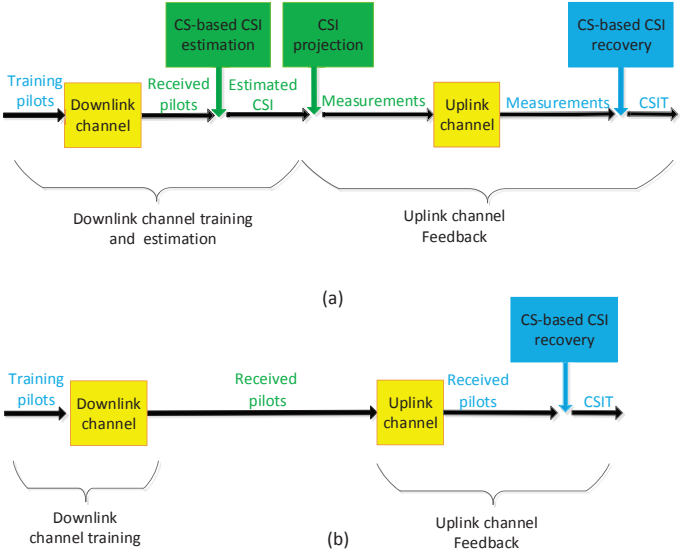}}
\caption{Comparison between conventional and proposed joint scheme: (a) Conventional CS-based channel training and feedback scheme. (b) Proposed CS-based joint channel training and feedback scheme.}
\label{TrainingFeedback}
\end{figure}

Conventional channel training and feedback schemes usually consists of two steps: downlink CSI estimation at user side and CSI feedback in the uplink.
Under the framework of CS, channel estimation algorithm \cite{ZhuXD,HanYJ} and channel feedback scheme \cite{protocals,IET} exploring the sparsity of the CSI have been proposed to achieve overhead reduction.
In these approaches, these two coupled communication procedures are optimized separately as depicted in Fig. \ref{TrainingFeedback} (a).


To address the drawbacks of the conventional schemes, we propose a novel CS-based joint scheme that considers channel training and feedback in a joint manner. The corresponding schematics is illustrated in Fig.\ref{TrainingFeedback} (b),
where users directly feed the received training pilots back to the BS without channel estimation, and then the BS recovers the CIR via CS algorithms.
Note that the proposed CS-based joint channel training and feedback scheme avoids the complex computations at power-limited users for channel estimation\footnote{The precondition is that CSI is not required at users, which is normal in massive MIMO systems with precoding \cite{ScalingupMIMO}. Otherwise, CSI can also be obtained by using the proposed differential operation and S-CoSaMP algorithm at users, which will achieve better estimation performance than conventional channel estimation scheme as will be shown in Section IV.} and CSI projection,
which not only relieves the computational burden for the user devices, but also brings channel feedback performance improvement, as will be verified by simulations in Section IV.
As the dedicated uplink channel can be modeled as an AWGN channel\footnote{In this paper, we discuss the analog feedback \cite{Samardzija_and_Mandayam,Caire_et_al,FeedbackChannel,GaoZ}, thus, the quantization noise of digital feedback is not considered.} with the same SNR of downlink as in \cite{FeedbackChannel}, the received pilots at the BS can still be modeled by (2) except that in this case, the noise parameter $\sigma_n^2$ denotes the overall noise power both in the downlink and uplink \cite{FeedbackChannel}.

\begin{figure}
\vspace{-4mm}
\centering{\includegraphics[width=0.5\textwidth]{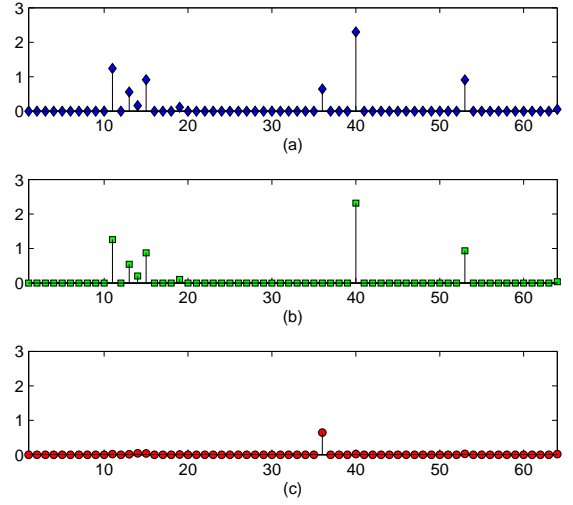}}
\vspace{-6mm}
\caption{A snapshot of the previous CIR, the current CIR, and the differential CIR: (a) the previous CIR; (b) the current CIR; (c) the differential CIR.}
\label{Diff_CIR}
\end{figure}

To exploit the temporal correlation of time-varying MIMO channels, we consider the \textit{differential} CIR between two CIRs in adjacent time slots, which can be expressed as
 \begin{align}
  \label{eq6}
  \Delta \mathbf{h}^{(t)}&\!=\!\mathbf{h}^{(t)}-\mathbf{h}^{(t-1)}\!\\\nonumber
  &\!=\!\mathbf{s}^{(t)}\!\circ\!(\mathbf{a}^{(t)}\!-\!\mathbf{a}^{(t-1)})\!+\!(\mathbf{s}^{(t)}\!-\!\mathbf{s}^{(t-1)})\!\circ\!\mathbf{a}^{(t-1)}\\\nonumber
  &\!=\!\mathbf{s}^{(t)}\!\circ\![\!\sqrt{1\!-\!\rho^2}\mathbf{w}\!\!-\!\!(1\!-\!\rho)\mathbf{a}^{(t\!-\!1)}]\!+\!(\mathbf{s}^{(t)}\!-\!\mathbf{s}^{(t-1)})\!\circ\!\mathbf{a}^{(t\!-\!1)}.
  \label{eq6}
\end{align}
When the movement velocity of users is not very high, e.g., $v=12$km/h, and the carrier frequency is 900MHz, the resulting Doppler frequency $f_d$ is 10Hz. For the typical time slot duration $\tau=0.5$ms, the correlation coefficient $\rho$ is 0.9911 \cite{protocals}.
Thus, the first term on the last line of (6) is close to zero.
On the other hand, since the delay indices of non-zero taps are varying slowly \cite{Dai14}, i.e., the CIR's support $\mathbf{s}^{(t)}$ changes slowly,
the number of non-zero elements of the second term is also small.
Fig. \ref{Diff_CIR} presents the snapshot of the previous CIR $\mathbf{h}^{(t-1)}$, the current CIR $\mathbf{h}^{(t)}$, and the differential CIR $\Delta \mathbf{h}^{(t)}$ of the channels described by (\ref{eq3}),  (\ref{eq4}) and (\ref{eq6}).
We can observe that the differential CIR enjoys a much stronger sparsity\footnote{The sparsity level $K'$ of the differential CIR is mainly dependent on the transition probability $p_{01}$ and $p_{10}$.} than the original CIR due to the temporal correlation of channels.

To utilize this observation, we modify (\ref{eq2}) and express the received pilots at the BS as
\begin{equation}
\mathbf{y}_\Omega^{(t)}=\mathbf{\Theta}\mathbf{h}^{(t)}+\mathbf{n}_\Omega^{(t)}.
\end{equation}
We try to exploit the temporal correlation of channels by computing the difference between received pilots at the BS in two adjacent time slots as
\begin{align}
\Delta\mathbf{y}_\Omega^{(t)}&=\mathbf{y}_\Omega^{(t)}-\mathbf{y}_\Omega^{(t-1)}\\\nonumber
&=\mathbf{\Theta}(\mathbf{h}^{(t)}-\mathbf{h}^{(t-1)})+\mathbf{n}_\Omega^{(t)}-\mathbf{n}_\Omega^{(t-1)}\\\nonumber
&=\mathbf{\Theta}\Delta\mathbf{h}^{(t)}+\Delta\mathbf{n}_\Omega^{(t)},
\end{align}
Where $\Delta \mathbf{h}^{(t)}=\mathbf{h}^{(t)}-\mathbf{h}^{(t-1)}$. As mentioned, the differential CIR $\Delta\mathbf{h}^{(t)}$ enjoys much stronger sparsity than the original CIR $\mathbf{h}^{(t)}$, which implies that better recovery performance by CS algorithm can be expected for the fixed pilot overhead, or equivalently, reduced pilot overhead can be achieved to obtain the same performance \cite{Stru_CS_Theory_Appli}.
After the differential CIR is recovered via CS algorithms, the current CIR can be obtained by adding the estimate of differential CIR and the estimate of the CIR in the previous slot.

Note that a precise CIR recovery in the initial time slot $\hat{\mathbf{h}}^{(0)}$ is important, since the recovery error in initial time slot can be propagated to the subsequent recovery process.
Additionally, an unexpected recovery error of the differential CIR will impair the subsequent CIR recovery process.
To avoid such error propagations, the proposed scheme will be re-initiated every $R$ time slots (either fixed or adaptive).
In the initial time slot, BS antennas transmit more pilots, denoted as $\mathbf{c}^0_i\in\mathcal{C}^{P^0\times1}$ for $i=1, 2,\cdots, M$,
where $P^0$ denotes the number of pilots in the initial time slot,
occupying more subcarriers, denoted as $\Omega^0$ and $\Omega^0\supset\Omega$.
Note that $\Omega^0\supset\Omega$ can be ensured by selecting a part of pilots in the initial time slots as pilots in the subsequent time slots.
It is well-known that the length of training sequence guaranteeing the reliable recovery of CSIT should be $\beta KM\ln (L/K)$ for $P$ and $\beta KM'\ln (L/K')$ for $P^0$, where $\beta$ is a constant scalar\footnote{The parameter $\beta$ can be usually set in the range from 1 to 5 for most practical applications \cite{CS_Eldar}.}, $K$ and $K'$ are the sparsity level of the original CIR and the differential CIR \cite{Stru_CS_Theory_Appli}, respectively.
Although such initialization will induce relatively large overhead in pilots, overall pilot overhead is not severe since the subsequent recovery of differential CIR will compensate it.
\subsection{S-CoSaMP Algorithm}

As discussed in Section II-C, $\mathbf{h}^{(t)}$ and $\mathbf{h}^{(t-1)}$ are both structured sparse, hence $\Delta\mathbf{h}^{(t)}$ is also structured sparse.
In this section, we present the S-CoSaMP algorithm exploiting this structured sparsity.
For simplicity, we omit the superscript $^{(t)}$ in this subsection.

Recalling the system model (2),
we aim to recover CIR $\mathbf{h}\in\mathcal{C}^{LM\times1}$ from the received pilots $\mathbf{y}_\Omega\in\mathcal{C}^{P\times1}$ at the BS.
For LS-based CIR recovery, the recovered CIR $\hat{\mathbf{h}}$ is given by
 \vspace{-1mm}
\begin{equation}
\hat{\mathbf{h}}=\mathbf{\Theta}^{\dagger}\mathbf{y}_\Omega,
 \vspace{-1mm}
\end{equation}
where $P\ge LM$ is required to ensure accurate recovery \cite{Ozdemir}.
This choice, obviously, causes prohibitive training and feedback overhead, in particular when $M$ is large.
When $P<LM$, (\ref{eq2}) becomes an underdetermined problem.
When the CIR vector is sparse, we can recover it using the CS technique.
In the CS theory, $\mathbf{y}_\Omega$ and $\mathbf{\Theta}$ are referred to as the measurements and the measurement (sensing) matrix, respectively.
\begin{algorithm}[htb]
\renewcommand{\algorithmicrequire}{\textbf{Input:}}
\renewcommand\algorithmicensure {\textbf{Output:} }
\caption{S-CoSaMP Algorithm}
\begin{algorithmic}[1]
\REQUIRE ~~
Received pilots $\mathbf{y}_\Omega$;
Measurement matrix $\mathbf{\Theta}$;
Sparsity level $K$.
\ENSURE ~~
CIR recovery $\mathbf{\hat{h}}$. \\
\STATE Initialization :
\STATE $\Gamma^0=\varnothing$, $\mathbf{r}=\mathbf{y}_\Omega$, $i=0$.
\WHILE {$i\leq 2KL$ and $||r||_2\geq\xi||\mathbf{y}_\Omega||_2$}
\STATE $i\leftarrow i+1$
\STATE $\mathbf{e}\leftarrow\mathbf{\Theta}^{H}\mathbf{r}$    \{form residual signal recovery\}
\STATE $\mathbf{z}(l)\leftarrow \sum_{m=0}^{M-1}\mathbf{e}(mL+l)$, $l=1,2,\cdots,L$
\STATE $\Lambda\leftarrow\Gamma_{\mathcal{T}(\mathbf{z},2K)}$   \{prune residual signal recovery\}
\STATE $\Lambda\leftarrow\Gamma^{i-1}\cup\Lambda$   \{merge support\}
\STATE $S\leftarrow\{mL+l\}$, $m=1,2,\cdots,M$, $l\in\Lambda$
\STATE $\mathbf{b}|_S\leftarrow\mathbf{\Theta}_S^\dagger\mathbf{y}_\Omega$, $\mathbf{b}|_{S^C}\leftarrow0$ \{form signal recovery\}
\STATE $\mathbf{g}(l)\leftarrow \sum_{m=0}^{m=M-1}\mathbf{b}(mL+l)$, $l=1,2,\cdots,L$
\STATE $\Gamma^i\leftarrow\Gamma_{\mathcal{T}(\mathbf{g},K)}$ \{prune signal recovery\}
\STATE $Q\leftarrow\{mL+l\}$, $m=1,2,\cdots,M$, $l\in\Gamma^i$
\STATE $\hat{\mathbf{h}}^i|_Q\leftarrow\mathbf{b}|_Q$, $\hat{\mathbf{h}}^i|_{Q^C}\leftarrow0$
\STATE $\mathbf{r}\leftarrow\mathbf{y}_\Omega-\mathbf{\Theta}\hat{\mathbf{h}}^i$  \{update measurement residual\}
\ENDWHILE
\RETURN $\mathbf{\hat{h}}=\mathbf{\hat{h}}^i$.
\end{algorithmic}
\end{algorithm}

We propose the S-CoSaMP algorithm to recover the CIR $\mathbf{h}^0$ in the initial time slot from the received pilots $\mathbf{y}_\Omega^{0}$ at the BS, as well as the differential CIR $\Delta \mathbf{h}^{(t)}$ from the differential pilots $\Delta\mathbf{y}_\Omega^{(t)}$ in the subsequent time slots.
The pseudocode of the proposed algorithm is provided in \textbf{Algorithm 1}\footnote{This algorithm is developed based on CoSaMP algorithm due to its low complexity and robustness.}.
Similar to the greedy CS algorithms such as orthogonal matching pursuit algorithm (OMP) \cite{Stru_CS_Theory_Appli},
we aim to find those columns of measurement matrix $\mathbf{\Theta}$ most correlated with the measurements $\mathbf{y}_\Omega$. After obtaining the column correlation (i.e., the residual signal estimate in step 5),
we keep 2$K$ columns of  $\mathbf{\Theta}$ most correlated to $\mathbf{y}_\Omega$, where $K$ is the signal sparsity level,
which can be merged with the support of signal estimated in previous iteration (step 8).
Then, we can estimate the signal by LS algorithm (step 10).
After that, the signal estimation will be pruned according to the sparsity level $K$ and the measurement residual can be updated for the next iteration (step 15).

The key idea of the proposed S-CoSaMP algorithm is that, the support of each of $\{\mathbf{h}_i\}_{i=1}^{M}$ is updated together, since $\mathbf{h}$ has the inherent structured sparsity.
Unlike the standard CoSaMP algorithm \cite{Stru_CS_Theory_Appli}, which does not consider the structured property of $\mathbf{h}$, the proposed S-CoSaMP algorithm offers a more precise support update by considering the structured sparsity of $\mathbf{h}$. In doing so, the CIR recovery performance can be improved, which can be verified by simulations in the next section.

\section{Simulation Results and Discussion}
This section investigates the performance of the proposed S-CoSaMP based differential joint channel training and feedback scheme.
The system setup is as follows:
1) The length of OFDM symbol is $N=2048$,
the number of BS antennas is $M=32$,
the maximal channel length is $L=64$;
2) The probability $\mu$ is set as 0.1, which means that the average channel sparsity level is $K=\mu L=6$,
the transition probability $p_{01}$ is set as 0.16,
the maximal Doppler frequency is $f_d=10$Hz,
the time slot duration is $\tau=0.5$ms, $\sigma_w^2=1$ and
the initial amplitudes $\mathbf{a}^{(0)}\thicksim\mathcal{CN}(0,1)$;
3) The proposed scheme will be initiated every $R=3$ time slots;
4) The coefficient $\xi$ in \textbf{Algorithm 1} is set as $10^{-3}$.
From equation (2), the signal-to-noise ratio (SNR) can be defined as $SNR=\frac{||\mathbf{\Theta h}||^2}{\sigma_n^2}$,
where the SNR is an overval SNR when $\sigma_n^2$ denotes the sum of noise power in the downlink and uplink.
For simplicity, we call it as SNR unless it causes confusion.
The same training overhead $\eta=P/N$ is considered to ensure fair comparision.
For the proposed differential scheme, the average pilot overhead is $\eta=\frac{P^0+P*(R-1)}{RN}$.

\begin{figure}
\center{\includegraphics[width=0.5\textwidth]{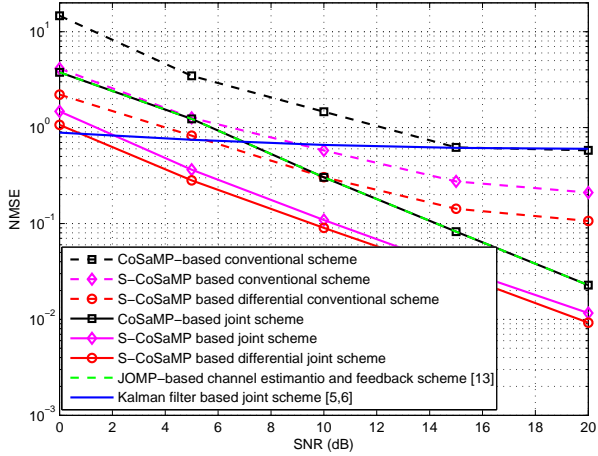}}
\vspace{-4mm}
\caption{NMSE performance comparison between the proposed joint scheme and the traditional scheme.}
\label{est_fb_JOMP_Kalman}
\end{figure}

Firstly, we compare the proposed joint channel training and feedback scheme with its conventional counterpart in Fig. \ref{est_fb_JOMP_Kalman}.
Note that the NMSE of joint orthogonal matching pursuit (J-OMP) algorithm \cite{RaoJ} and Kalman filter \cite{Noh_et_al,Choi_et_al} are also presented for comparison.
The dotted lines denote the conventional scheme.
Note that the conventional scheme treats channel training and feedback separately.
We observe that the proposed joint scheme outperforms the conventional scheme, which means the joint consideration of channel training and feedback improves the CIR recovery performance at the BS.
In addition, S-CoSaMP based differential joint scheme achieves the best NMSE performance due to the exploiting of the temporal correlation and structured sparsity.
The performance of J-OMP and CoSaMP are similar because the channel assumption of common support is not satisfied and J-OMP degrades as OMP, which has similar recovery performance with CoSaMP.
The performance of Kalman filter is not good when the channel autocorrelation information is unknown \cite{Noh_et_al}.

\begin{figure}
\center{\includegraphics[width=0.5\textwidth]{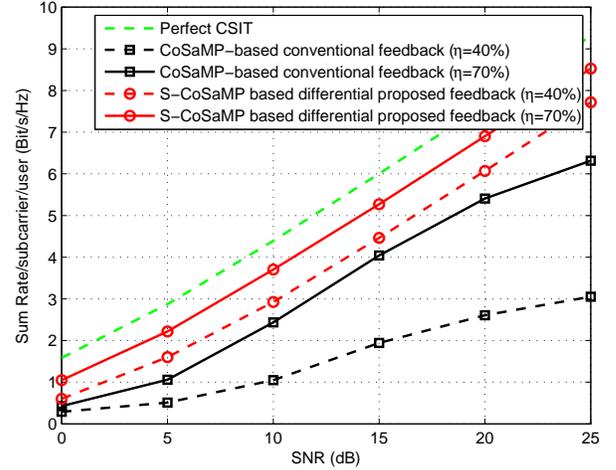}}
\vspace{-4mm}
\caption{Downlink data rate performance comparison with two different training and feedback overhead: $\eta=40\%$ and $\eta=70\%$.}
\label{est_fb_non_lock_in_Rs}
\end{figure}
Fig. \ref{est_fb_non_lock_in_Rs} shows the data rate comparison when two different channel training and feedback overheads are considered: $\eta=40\%$ and $\eta=70\%$. The ideal case where the CSIT is perfectly known at the BS is also presented as a benchmark for comparison.
As shown in Fig. \ref{est_fb_non_lock_in_Rs}, S-CoSaMP based differential joint scheme outperforms CoSaMP-based conventional scheme in both cases.
Note that, S-CoSaMP based differential joint scheme with $\eta=40\%$ and CoSaMP-based conventional scheme with $\eta=70\%$ can achieve similar data rate.
This means that the proposed scheme achieves around $30\%$ reduction in channel training and feedback overhead over the conventional scheme in achieving the same data rate.

\section{Conclusion}
In this paper, we have investigated the challenging problem of channel training and feedback for FDD massive MIMO systems.
By exploiting the temporal correlation of MIMO channels, we have proposed the differential joint channel training and feedback scheme, where users directly feed back the received pilots and then the explicit CSI can be obtained by CS algorithms at the BS.
By exploiting the structured sparsity of MIMO channels, we have proposed the S-CoSaMP algorithm to further reduce the overhead.
In the future, we will consider the spatial correlation of CSI from different users that is available at the BS to further reduce the overhead.

\end{document}